\input harvmac
\input amssym
\input epsf

\def\unit{\relax{\rm 1\kern-.26em I}}
\def\nada{\relax{\rm 0\kern-.30em l}}
\def\tilde{\widetilde}



\def\IL{\relax{\rm I\kern-.18em L}}
\def\IH{\relax{\rm I\kern-.18em H}}
\def\IR{\relax{\rm I\kern-.18em R}}
\def\IC{\relax\hbox{$\inbar\kern-.3em{\rm C}$}}
\def\IZ{\relax\ifmmode\mathchoice
{\hbox{\cmss Z\kern-.4em Z}}{\hbox{\cmss Z\kern-.4em Z}}
{\lower.9pt\hbox{\cmsss Z\kern-.4em Z}} {\lower1.2pt\hbox{\cmsss
Z\kern-.4em Z}}\else{\cmss Z\kern-.4em Z}\fi}

\def\CJ {{\cal J}}

\def\CL {{\cal L}}
\def\CV {{\cal V}}
\def\CO {{\cal O}}

\def\CG {{\cal G}}
\def\CH {{\cal H}}


\def\CO {{\cal O}}

\def\CV{{\cal V }}

\def\Tr{{\rm Tr}}

\font\manual=manfnt \def\dbend{\lower3.5pt\hbox{\manual\char127}}

\def\IZ{\relax\ifmmode\mathchoice
{\hbox{\cmss Z\kern-.4em Z}}{\hbox{\cmss Z\kern-.4em Z}}
{\lower.9pt\hbox{\cmsss Z\kern-.4em Z}} {\lower1.2pt\hbox{\cmsss
Z\kern-.4em Z}}\else{\cmss Z\kern-.4em Z}\fi}

\def\lfm#1{\medskip\noindent\item{#1}}

\def\bar{\overline}

\def\CH{{\cal H}}

\def\rt2{\sqrt{2}}
\def\irt2{{1\over\sqrt{2}}}

\def\hat{\widehat}
\def\slashchar#1{\setbox0=\hbox{$#1$}           
  \dimen0=\wd0                                 
  \setbox1=\hbox{/} \dimen1=\wd1               
  \ifdim\dimen0>\dimen1                        
     \rlap{\hbox to \dimen0{\hfil/\hfil}}      
     #1                                        
  \else                                        
     \rlap{\hbox to \dimen1{\hfil$#1$\hfil}}   
     /                                         
  \fi}

\def\foursqr#1#2{{\vcenter{\vbox{
   \hrule height.#2pt
   \hbox{\vrule width.#2pt height#1pt \kern#1pt
   \vrule width.#2pt}
   \hrule height.#2pt
   \hrule height.#2pt
   \hbox{\vrule width.#2pt height#1pt \kern#1pt
   \vrule width.#2pt}
   \hrule height.#2pt
       \hrule height.#2pt
   \hbox{\vrule width.#2pt height#1pt \kern#1pt
   \vrule width.#2pt}
   \hrule height.#2pt
       \hrule height.#2pt
   \hbox{\vrule width.#2pt height#1pt \kern#1pt
   \vrule width.#2pt}
   \hrule height.#2pt}}}}
\def\psqr#1#2{{\vcenter{\vbox{\hrule height.#2pt
   \hbox{\vrule width.#2pt height#1pt \kern#1pt
   \vrule width.#2pt}
   \hrule height.#2pt \hrule height.#2pt
   \hbox{\vrule width.#2pt height#1pt \kern#1pt
   \vrule width.#2pt}
   \hrule height.#2pt}}}}
\def\sqr#1#2{{\vcenter{\vbox{\hrule height.#2pt
   \hbox{\vrule width.#2pt height#1pt \kern#1pt
   \vrule width.#2pt}
   \hrule height.#2pt}}}}

\def\figin{\epsfcheck\figin}\def\figins{\epsfcheck\figins}
\def\epsfcheck{\ifx\epsfbox\UnDeFiNeD
\message{(NO epsf.tex, FIGURES WILL BE IGNORED)}
\gdef\figin##1{\vskip2in}\gdef\figins##1{\hskip.5in}
\else\message{(FIGURES WILL BE INCLUDED)}%
\gdef\figin##1{##1}\gdef\figins##1{##1}\fi}
\def\DefWarn#1{}
\def\figinsert{\goodbreak\midinsert}
\def\ifig#1#2#3{\DefWarn#1\xdef#1{fig.~\the\figno}
\writedef{#1\leftbracket fig.\noexpand~\the\figno}%
\figinsert\figin{\centerline{#3}}\medskip\centerline{\vbox{\baselineskip12pt
\advance\hsize by -1truein\noindent\footnotefont{\bf
Fig.~\the\figno:\ } \it#2}}
\bigskip\endinsert\global\advance\figno by1}

\lref\gmreview{
 G.~F.~Giudice and R.~Rattazzi,
 ``Theories with gauge-mediated supersymmetry breaking,''
 Phys.\ Rept.\  {\bf 322}, 419 (1999)
 [arXiv:hep-ph/9801271].
}

\lref\AffleckXZ{
 I.~Affleck, M.~Dine and N.~Seiberg,
 ``Dynamical Supersymmetry Breaking In Four-Dimensions And Its
 Phenomenological Implications,''
 Nucl.\ Phys.\  B {\bf 256}, 557 (1985).
}

\lref\DineYW{
 M.~Dine and A.~E.~Nelson,
 ``Dynamical supersymmetry breaking at low-energies,''
 Phys.\ Rev.\  D {\bf 48}, 1277 (1993)
 [arXiv:hep-ph/9303230].
}

\lref\DineVC{
 M.~Dine, A.~E.~Nelson and Y.~Shirman,
 ``Low-Energy Dynamical Supersymmetry Breaking Simplified,''
 Phys.\ Rev.\  D {\bf 51}, 1362 (1995)
 [arXiv:hep-ph/9408384].
}

\lref\LutyFK{
 M.~A.~Luty,
 ``Naive dimensional analysis and supersymmetry,''
 Phys.\ Rev.\  D {\bf 57}, 1531 (1998)
 [arXiv:hep-ph/9706235].
}

\lref\DineAG{
 M.~Dine, A.~E.~Nelson, Y.~Nir and Y.~Shirman,
 ``New tools for low-energy dynamical supersymmetry breaking,''
 Phys.\ Rev.\  D {\bf 53}, 2658 (1996)
 [arXiv:hep-ph/9507378].
}

\lref\WittenKV{
 E.~Witten,
 ``Mass Hierarchies In Supersymmetric Theories,''
 Phys.\ Lett.\  B {\bf 105}, 267 (1981).
}

\lref\DermisekQJ{
 R.~Dermisek, H.~D.~Kim and I.~W.~Kim,
 ``Mediation of supersymmetry breaking in gauge messenger models,''
 JHEP {\bf 0610}, 001 (2006)
 [arXiv:hep-ph/0607169].
}

\lref\DineGU{
 M.~Dine and W.~Fischler,
 ``A Phenomenological Model Of Particle Physics Based On Supersymmetry,''
 Phys.\ Lett.\  B {\bf 110}, 227 (1982).
}

\lref\NappiHM{
 C.~R.~Nappi and B.~A.~Ovrut,
 ``Supersymmetric Extension Of The SU(3) X SU(2) X U(1) Model,''
 Phys.\ Lett.\  B {\bf 113}, 175 (1982).
}

\lref\DineZB{
 M.~Dine and W.~Fischler,
 ``A Supersymmetric Gut,''
 Nucl.\ Phys.\  B {\bf 204}, 346 (1982).
}

\lref\AlvarezGaumeWY{
 L.~Alvarez-Gaume, M.~Claudson and M.~B.~Wise,
 ``Low-Energy Supersymmetry,''
 Nucl.\ Phys.\  B {\bf 207}, 96 (1982).
}

\lref\dimgiud{
S.~Dimopoulos and G.~F.~Giudice,
 ``Multi-messenger theories of gauge-mediated supersymmetry breaking,''
 Phys.\ Lett.\  B {\bf 393}, 72 (1997)
 [arXiv:hep-ph/9609344].
}


\lref\kawamura{
 Y.~Kawamura, H.~Murayama and M.~Yamaguchi,
 ``Probing symmetry breaking pattern using sfermion masses,''
 Phys.\ Lett.\  B {\bf 324}, 52 (1994)
 [arXiv:hep-ph/9402254].
}


\lref\MartinZB{
 S.~P.~Martin,
 ``Generalized messengers of supersymmetry breaking and the sparticle mass
 spectrum,''
 Phys.\ Rev.\  D {\bf 55}, 3177 (1997)
 [arXiv:hep-ph/9608224].
}

\lref\KomargodskiAX{
 Z.~Komargodski and N.~Seiberg,
 ``mu and General Gauge Mediation,''
 JHEP {\bf 0903}, 072 (2009)
 [arXiv:0812.3900 [hep-ph]].
}
\lref\MasonIQ{
 J.~D.~Mason,
 ``Gauge Mediation with a small mu term and light squarks,''
 arXiv:0904.4485 [hep-ph].
}

\lref\CsakiSR{
 C.~Csaki, A.~Falkowski, Y.~Nomura and T.~Volansky,
 ``New Approach to the mu-Bmu Problem of Gauge-Mediated Supersymmetry
 Breaking,''
 Phys.\ Rev.\ Lett.\  {\bf 102}, 111801 (2009)
 [arXiv:0809.4492 [hep-ph]].
}

\lref\IntriligatorFR{
 K.~A.~Intriligator and M.~Sudano,
 ``Comments on General Gauge Mediation,''
 JHEP {\bf 0811}, 008 (2008)
 [arXiv:0807.3942 [hep-ph]].
}

\lref\GorbatovQA{
 E.~Gorbatov and M.~Sudano,
 ``Sparticle Masses in Higgsed Gauge Mediation,''
 JHEP {\bf 0810}, 066 (2008)
 [arXiv:0802.0555 [hep-ph]].
}

\lref\LuoKF{
 M.~Luo and S.~Zheng,
 ``Gauge Extensions of Supersymmetric Models and Hidden Valleys,''
 JHEP {\bf 0904}, 122 (2009)
 [arXiv:0901.2613 [hep-ph]].
}

\lref\MeadeWD{
 P.~Meade, N.~Seiberg and D.~Shih,
 ``General Gauge Mediation,''
 Prog.\ Theor.\ Phys.\ Suppl.\  {\bf 177}, 143 (2009)
 [arXiv:0801.3278 [hep-ph]].
}

\lref\BuicanWS{
 M.~Buican, P.~Meade, N.~Seiberg and D.~Shih,
 ``Exploring General Gauge Mediation,''
 JHEP {\bf 0903}, 016 (2009)
 [arXiv:0812.3668 [hep-ph]].
}

\lref\KomargodskiJF{
 Z.~Komargodski and D.~Shih,
 ``Notes on SUSY and R-Symmetry Breaking in Wess-Zumino Models,''
 JHEP {\bf 0904}, 093 (2009)
 [arXiv:0902.0030 [hep-th]].
}

\lref\RayWK{
 S.~Ray,
 ``Some properties of meta-stable supersymmetry-breaking vacua in Wess-Zumino
 models,''
 Phys.\ Lett.\  B {\bf 642}, 137 (2006)
 [arXiv:hep-th/0607172].
}

\lref\LangackerYV{
 P.~Langacker,
 ``The Physics of Heavy $Z^\prime$ Gauge Bosons,''
 arXiv:0801.1345 [hep-ph].
}

\lref\DimopoulosIG{
 S.~Dimopoulos and G.~F.~Giudice,
 ``Multi-messenger theories of gauge-mediated supersymmetry breaking,''
 Phys.\ Lett.\  B {\bf 393}, 72 (1997)
 [arXiv:hep-ph/9609344].
}

\lref\FoxBU{
 P.~J.~Fox, A.~E.~Nelson and N.~Weiner,
 ``Dirac gaugino masses and supersoft supersymmetry breaking,''
 JHEP {\bf 0208}, 035 (2002)
 [arXiv:hep-ph/0206096].
}

\lref\AmigoRC{
 S.~D.~L.~Amigo, A.~E.~Blechman, P.~J.~Fox and E.~Poppitz,
 ``R-symmetric gauge mediation,''
 JHEP {\bf 0901}, 018 (2009)
 [arXiv:0809.1112 [hep-ph]].
}

\lref\BenakliPG{
 K.~Benakli and M.~D.~Goodsell,
 ``Dirac Gauginos in General Gauge Mediation,''
 Nucl.\ Phys.\  B {\bf 816}, 185 (2009)
 [arXiv:0811.4409 [hep-ph]].
}

\lref\GiudiceBP{
 G.~F.~Giudice and R.~Rattazzi,
 ``Theories with gauge-mediated supersymmetry breaking,''
 Phys.\ Rept.\  {\bf 322}, 419 (1999)
 [arXiv:hep-ph/9801271].
}

\lref\WittenNF{
 E.~Witten,
 ``Dynamical Breaking Of Supersymmetry,''
 Nucl.\ Phys.\  B {\bf 188}, 513 (1981).
}

\lref\DineZA{
 M.~Dine, W.~Fischler and M.~Srednicki,
``Supersymmetric Technicolor,''
 Nucl.\ Phys.\  B {\bf 189}, 575 (1981).
}

\lref\DimopoulosAU{
 S.~Dimopoulos and S.~Raby,
``Supercolor,''
 Nucl.\ Phys.\  B {\bf 192}, 353 (1981).
}

\lref\DineGU{
 M.~Dine and W.~Fischler,
``A Phenomenological Model Of Particle Physics Based On Supersymmetry,''
 Phys.\ Lett.\  B {\bf 110}, 227 (1982).
}

\lref\NappiHM{
 C.~R.~Nappi and B.~A.~Ovrut,
 ``Supersymmetric Extension Of The SU(3) X SU(2) X U(1) Model,''
 Phys.\ Lett.\  B {\bf 113}, 175 (1982).
}

\lref\AlvarezGaumeWY{
 L.~Alvarez-Gaume, M.~Claudson and M.~B.~Wise,
``Low-Energy Supersymmetry,''
 Nucl.\ Phys.\  B {\bf 207}, 96 (1982).
}

\lref\DimopoulosGM{
 S.~Dimopoulos and S.~Raby,
 ``Geometric Hierarchy,''
 Nucl.\ Phys.\  B {\bf 219}, 479 (1983).
}

\lref\LangackerIP{
  P.~Langacker, G.~Paz, L.~T.~Wang and I.~Yavin,
  ``Aspects of Z'-mediated Supersymmetry Breaking,''
  Phys.\ Rev.\  D {\bf 77}, 085033 (2008)
  [arXiv:0801.3693 [hep-ph]].
}

\lref\LangackerAC{
  P.~Langacker, G.~Paz, L.~T.~Wang and I.~Yavin,
  ``Z'-mediated Supersymmetry Breaking,''
  Phys.\ Rev.\ Lett.\  {\bf 100}, 041802 (2008)
  [arXiv:0710.1632 [hep-ph]].
}

\lref\PoppitzXW{
  E.~Poppitz and S.~P.~Trivedi,
  ``Some remarks on gauge-mediated supersymmetry breaking,''
  Phys.\ Lett.\  B {\bf 401}, 38 (1997)
  [arXiv:hep-ph/9703246].
}

\lref\NelsonJI{
  A.~E.~Nelson and N.~J.~Weiner,
  ``Gauge/anomaly Syzygy and generalized brane world models of  supersymmetry
  breaking,''
  Phys.\ Rev.\ Lett.\  {\bf 88}, 231802 (2002)
  [arXiv:hep-ph/0112210].
}

\lref\CarpenterRJ{
  L.~M.~Carpenter,
  ``Gauge Mediation with D-terms,''
  arXiv:0809.0026 [hep-ph].
}

\lref\IntriligatorBE{
  K.~Intriligator and M.~Sudano,
  ``General Gauge Mediation with Gauge Messengers,''
  arXiv:1001.5443 [hep-ph].
}

\rightline{CERN-PH-TH/2009-178}
\Title{\vbox{\baselineskip12pt }} {\vbox{\centerline{ Soft Terms from Broken Symmetries}}}
\smallskip
\centerline{Matthew Buican$^\dagger$ and Zohar Komargodski$^*$}
\smallskip
\bigskip
\centerline{$^\dagger${\it Department of Physics, CERN Theory Division CH-1211, Geneva 23, Switzerland}} \centerline{$^*${\it School
of Natural Sciences, Institute for Advanced Study, Princeton, NJ
08540 USA}} \vskip 1cm

\noindent In theories of phyiscs beyond the Standard Model (SM), visible sector fields often carry quantum numbers under additional gauge symmetries. One could then imagine a scenario in which these extra gauge symmetries play a role in transmitting supersymmetry breaking from a hidden sector to the Supersymmetric Standard Model (SSM). In this paper we present a general formalism for studying the resulting hidden sectors and calculating the corresponding gauge mediated soft parameters. We find that a large class of generic models features a leading universal contribution to the soft scalar masses that only depends on the scale of Higgsing, even if the model is strongly coupled. As a by-product of our analysis, we elucidate some IR aspects of the correlation functions in General Gauge Mediation. We also discuss possible phenomenological applications.
\bigskip
\Date{September 2009}


\newsec{Introduction}
Low-scale supersymmetry (SUSY) breaking offers a viable and attractive scenario for physics beyond the Standard Model (SM). However, the specific details of the new physics are model-dependent. A particularly well-motivated approach is to consider field-theoretical supersymmetry breaking and its gauge mediation~\refs{\DineZA\DimopoulosAU\DineGU\NappiHM\AlvarezGaumeWY-\DimopoulosGM} to the visible sector. From the theoretical standpoint it is well defined and often calculable. It is also appealing phenomenologically, as  it automatically  addresses the flavor puzzle and in addition may help to elucidate the hierarchy between the weak scale and the Planck scale~\WittenNF.

Armed with these motivations, one is driven to study the predictions of gauge mediation in more detail. To that end, the authors of \MeadeWD\ gave a general definition of gauge mediation. The basic idea, dubbed \lq\lq General Gauge Mediation" (GGM), is to define gauge mediation as a scenario in which a theory splits into a sector consisting of a supersymmetric extension of the Standard Model (SSM) and a separate, decoupled, SUSY-breaking hidden sector, $\CH$, as one takes the various SSM gauge couplings $g_r\to0$. The labels $r=1,2,3$ represent the $U(1)_Y$, $SU(2)_W$, and $SU(3)_C$ factors of the SSM gauge group respectively.

Even under this broad definition, the authors of~\MeadeWD\ found that the soft scalar masses arising in theories of gauge mediation obey two sum rules\foot{The masses are then subject to renormalization group evolution from the scale, $M$, at which the sum rules are defined down to the infrared.}
\eqn\twosumrules{\Tr\left(Ym^2\right)=0~,\qquad \Tr\left((B-L)m^2\right)=0~.}
In fact, one can show that the two sum rules are the only constraints generically obeyed by the scalar masses~\BuicanWS.

One can generalize these discussions from the case of pure gauge mediation to the case in which there are also direct couplings between the Higgs sector of the SSM and $\CH$. Studying these couplings, one can deduce various features of the soft parameters (for some recent works see~\refs{\CsakiSR,\KomargodskiAX,\MasonIQ}).  In many cases, however, the sum rules~\twosumrules\ for the scalar masses are essentially unaffected by the details of the Higgs couplings. Given this picture, it is tempting to conclude that the scalar sum rules may constitute a proverbial \lq\lq smoking gun" for gauge mediation.

However, our discussion thus far has been rooted in the assumption that the SSM matter fields are only charged under $SU(3)_C\times SU(2)_W\times U(1)_Y$. There are a few reasons to study a specific extension of this ansatz. Many constructions in string theory and field theory lead to scenarios in which the SSM matter fields are charged under additional gauge groups which are Higgsed above the electroweak scale. These could be various different $U(1)$ gauge symmetries or the more conventional $U(1)_{B-L}$. The literature on this subject is vast, see the recent review~\LangackerYV\ for details and references. The scale of breaking of these additional symmetries depends upon the assumptions of the model and can be in a range of energies from the electroweak scale to the GUT scale. In this paper we investigate the effects of such Higgsed symmetries on gauge mediation. We will see that the inclusion of Higgsed symmetries leads to changes in the scalar masses and (partial or complete) violation of the sum rules. On the other hand, we will show that the inclusion of additional Higgsed symmetries leads to certain universal predictions.

In order to proceed with our discussion, we first generalize the definition of gauge mediation given above to accommodate the presence of additional, spontaneously broken, symmetries. To that end, let us define
                             \eqn\totalgroup{\CG_{SSM}\supset SU(3)_C\times SU(2)_W\times U(1)_Y}
to be the gauge group under which the SSM fields are charged. Then, we define gauge mediation to be a scenario in which our theory splits into the SSM and a decoupled SUSY-breaking hidden sector, $\CH$, whose vacuum spontaneously breaks \eqn\breaking{\CG_{SSM}\to SU(3)_C\times SU(2)_W\times U(1)_Y} in the limit that we take the gauge couplings of $\CG_{SSM}$ to zero. This definition implies that $\CG_{SSM}$ can be embedded in the global symmetry group, $\CG$, of~$\CH$.

A prototype of the scenario we are interested in is when the symmetry group is ${\cal G}_{SSM}=SU(3)_C\times SU(2)_W\times U(1)_Y\times U(1)$. Our goal is to calculate the observable soft masses in this setup as well as to uncover the structure of the hidden sector. Certain aspects of this problem have been discussed before in some limiting cases~\refs{\GorbatovQA,\IntriligatorFR,\LuoKF}. Here we will try to be as general as possible and to encompass a large class of models.

Since a main theme here is the spontaneous breaking of global symmetries (and thus the presence of massless particles), one has to have control over the IR behavior of various current correlation functions in $\CH$. In section~2 we explore the IR properties of these correlation functions. It turns out that this discussion is also relevant to the original setup of GGM, since there is always at least one massless particle when SUSY is broken (the Goldstino), and one has to make sure that it does not render the observable scalar masses IR sensitive. In particular, for models of SUSY breaking that do not have a \lq\lq messenger parity" symmetry, we provide a direct argument for the IR safety of the visible sector observables.\foot{Models without exact messenger parity are often viable, see e.g.~\DimopoulosIG.}

In section~3 we further define and discuss the general setup we are interested in and explain its predictions. We identify the scales at which various aspects of the dynamics take place, and we find that the leading contribution to the soft scalar masses in these theories is under full control even when the hidden sector is not perturbative. The crucial point is that the leading contribution to the scalar masses is controlled by the universal IR properties of the correlation functions we discuss in section 2. This situation is unlike the usual case of gauge mediation, where the scalar masses are often incalculable. In other words, to calculate the scalar masses in the usual scenario of gauge mediation one has to have control over energy ranges where intricate dynamics may take place. In our case we find that there are contributions from various energy scales, but the dominant contribution (in a systematic expansion in the gauge coupling) comes from the deep IR.

The result is that by expanding the scalar masses in the coupling, $g$, of the additional $U(1)$ symmetry, we find that the leading term is
\eqn\universalintro{\delta m^2_{soft}={3q^2g^4\over 16\pi^2}\log(g^2)f_\pi^2~,}
where $f_\pi$ is the decay constant of the pion of the spontaneously broken $U(1)$ and $q$ is the $U(1)$ charge of the sparticle of interest. Note that \universalintro\ contains a logarithm of the gauge coupling and is therefore conceptually different from the usual gauge mediation scenario. Furthermore, the contribution~\universalintro\ is negative and, as a result, can be used to lower the soft scalar masses. This is phenomenologically desirable. We will explain the origin of this universal result in detail and comment on the form of  subleading model dependent corrections to~\universalintro.\foot{Another important---though fundamentally different---logarithmic contribution to the scalar masses arises in theories that have non-vanishing messenger supertrace \PoppitzXW, \BuicanWS. Some phenomenological consequences of this effect have been discussed, for example, in \NelsonJI\  and \CarpenterRJ.}

A single additional (Higgsed) $U(1)$ symmetry always retains one of
the two sum rules~\twosumrules.\foot{Strictly speaking, this
statement is true when the mixed contributions from the two $U(1)$
factors are small. This could be due to exact or approximate
messenger parity of either of the $U(1)$ factors, or the embedding
of either of them in some non-Abelian group.} In the case that we
identify $U(1)=U(1)_{B-L}$, the first sum rule is still satisfied,
but the second sum rule is violated. Nevertheless, we know exactly
how it is violated at leading order in $g$. For example,  we can
predict that \eqn\pred{\Tr\left((B-L)m^2\right)>0~.}

In section~4 we describe a toy example where many of the ideas we discuss are manifest. In section~5 we conclude with a discussion and comments on possible phenomenological applications.

\bigskip
\noindent {\it Note: While completing this project, we learned of closely related work by K.~Intriligator and M.~Sudano~\IntriligatorBE.}

\newsec{The Supermultiplet of Global Symmetries}
For the purposes of this section, we will imagine that our hidden sector, $\CH$, has a single characteristic scale, $M$. This assumption is made for simplicity and can be relaxed straightforwardly.
Furthermore, we specialize to the case that $\CH$ is endowed with a global symmetry $\CG=U(1)$ and study the correlation functions of the associated symmetry current superfield, $\CJ$.

To begin our study, let us recall that the conserved $U(1)$ current, $j_{\mu}$, is packaged in a current superfield, $\CJ$, that satisfies the SUSY generalization of current conservation
\eqn\currcond{
D^2\CJ=0~.
}
This condition in turn implies that $\CJ$ can be expanded in superspace as
\eqn\currcomp{
\CJ=J+i\theta j-i\bar\theta\ \bar j-\theta\sigma^{\mu}\bar\theta j_{\mu}+{1\over2}\theta^2\bar\theta\bar\sigma^{\mu}\partial_{\mu}j-{1\over2}\bar\theta^2\theta\sigma^{\mu}\partial_{\mu}\bar j-{1\over4}\theta^2\bar\theta^2\partial^2 J~,
}
where $j_{\mu}$ is conserved, i.e., $\partial^{\mu}j_{\mu}=0$. From \currcomp\ one can read off all the SUSY transformations of the component fields. For instance, we see that
\eqn\CJsusytransf{\eqalign{
& \delta_\alpha j_\beta=0~,\qquad  \bar \delta_{\dot\alpha} \bar j_{\dot\beta} = 0~.
}}
Notice that our discussion so far accommodates the case in which the $U(1)$ is Higgsed, since \currcond\ is an operator equation and is therefore independent of the particular vacuum we find ourselves in.

\subsec{Correlation Functions} Let us now consider the simplest Euclidean correlation functions. The constraints
we impose are consistency under the Euclidean isometry group and current conservation. The only allowed
one-point function is \eqn\onept{ \langle J\rangle=\zeta~.} Some theories have an unbroken messenger parity
symmetry under which $\CJ\rightarrow-\CJ$. In this case we see that $\zeta=0$. Here we do not assume unbroken
messenger parity, although it will play an important role in our phenomenological discussions later.

Additionally, we find that there are, in principle, {\it five} allowed two-point functions
\eqn\currtwopt{\eqalign{
&\langle J(x)J(0)\rangle ={1\over x^4} C_0(x^2M^2)~,\cr&\langle j_{\alpha}(x)\bar j_{\dot\alpha}(0)\rangle=-i\sigma^{\mu}_{\alpha\dot\alpha}\partial_{\mu}\left({1\over x^4}C_{1/2}(x^2M^2)\right)~,\cr&\langle j_{\mu}(x)j_{\nu}(0)\rangle=(\eta_{\mu\nu}\partial^2-\partial_{\mu}\partial_{\nu})\left({1\over x^4} C_1(x^2M^2)\right)~,\cr&\langle j_{\alpha}(x)j_{\beta}(0)\rangle=\epsilon_{\alpha\beta}{1\over x^5}B_{1/2}(x^2M^2)~,\cr&\langle j_{\mu}(x)J(0)\rangle =\ cM^2 \ \partial_{\mu}\left({1\over x^2}\right)~,
}}
where $M$, as mentioned above, is the typical scale of the hidden sector.

The last correlation function, $\langle j_{\mu}(x)J(0)\rangle$, is special because its functional dependence is fixed. From the long distance power-law behavior of the $\langle j_{\mu}(x)J(0)\rangle$ correlator, we see that it must arise from the exchange of a massless boson. In particular, this massless boson should be created from the vacuum by acting with $j_\mu$. If the $U(1)$ symmetry is not Higgsed this cannot occur and therefore we conclude that in this case $c=0$, as mentioned in~\MeadeWD.

The more interesting situation is when the symmetry is Higgsed and a pion is created from the vacuum by acting with $j_\mu$.
It turns out that for the case we discuss here, $\CG=U(1)$, $\langle j_{\mu}(x)J(0)\rangle$ still vanishes as we shall now explain.
The formal proof proceeds as follows. From the last line of~\currtwopt\ we see that \eqn\twopoint{\eqalign{
&\langle j_{0}(x)J(0)\rangle\sim cM^2{x_0\over x^4}~.}} To calculate the commutator from this Euclidean
correlation function we integrate over $x_0=\epsilon$ and subtract the integral over $x_0=-\epsilon$. We get the
equal time commutator in Minkowski signature \eqn\comm{\biggl\langle \bigl[\int d^3x
j_0(\vec{x}),J(\vec{0})\bigr]\biggr\rangle\biggl|_{x_0=0}\sim cM^2~.} The fact that $J$ is neutral under the
$U(1)$ charge implies that $c=0$ and that, therefore, the last correlation function in~\currtwopt\ is zero. This
proof goes through as long as the group is Abelian.

Unlike the $\langle j_{\mu}(x)J(0)\rangle$ correlation function, the first four correlators in~\currtwopt\ are in general non-vanishing. In the presence of massless particles, their long-distance behavior is governed by the exchanges of these particles.

Generally speaking, the massless particles fall into one of two categories:
\lfm{\bf 1.} Particles associated with spontaneously broken symmetries of the hidden sector, i.e., the goldstino and the Goldstone boson(s) (including, possibly, an R-axion).

\lfm{\bf 2.} Massless fermions required for anomaly matching. We will sometimes refer to these particles as 't Hooft fermions.

\noindent
Of course, in theories with various sectors decoupled from the SUSY breaking, there could be additional massless particles not of the type above. This rather baroque possibility will be ignored here.

Our goal is to elucidate the IR properties of the four nontrivial correlation functions in~\currtwopt. To accomplish this let us first study the IR behavior of correlation functions involving the supercurrent. The reason we study these correlators is that they will be helpful in understanding the effects of the Goldstino.

Recall that the supercurrent $S_{\mu\alpha}$ is conserved, $\partial^\mu S_{\mu \alpha}=0$, and that in the deep low-energy regime it becomes the Goldstino, i.e. $S_{\mu\alpha}\sim F\sigma_{\mu \alpha\dot\alpha} \bar G^{\dot\alpha}$. Indeed, the spin $3/2$ component of the supercurrent decouples from the low energy physics (up to a possible improvement term) and therefore can be ignored in our analysis. We will be interested in the following correlation functions
\eqn\threeid{\eqalign{
& \langle \bar S^{\mu}_{\dot\alpha}(x)j_{\beta}(0)\rangle=\sigma^{\mu}_{\beta\dot\alpha}\partial^2 g(x^2M^2)-\sigma^{\nu}_{\beta\dot\alpha}\partial_{\nu}\partial^{\mu}g(x^2M^2)~,
\cr&\langle S^{\mu}_{\alpha}(x)j_{\beta}(0)\rangle=\tilde c\epsilon_{\alpha\beta}\partial^{\mu}\left({1\over x^2}\right)+\sigma^{\mu\nu}_{\alpha\beta}\partial_{\nu}\tilde g(x^2M^2)~,}}
where we have written the most generally allowed decomposition of these correlation functions consistent with current conservation.

We can further restrict the
form of the second correlator in~\threeid\ by repeating the logic around~\comm. Due to~\CJsusytransf\ and the
fact that the term proportional to $\tilde c$ leads to a nonzero commutator of the supercharge with $j_\beta$ we
conclude that $\tilde c=0$. Hence, \eqn\Sjii{ \langle
S^{\mu}_{\alpha}(x)j_{\beta}(0)\rangle=\sigma^{\mu\nu}_{\alpha\beta}\partial_{\nu}\tilde g(x^2M^2)~. } In order, to say
something about the unknown functions $g,\tilde g$ we use the fact that at large separation the supercurrent
becomes the Goldstino and thus \eqn\threeidIR{\eqalign{ &\lim_{|x|\rightarrow\infty}\langle \bar
S^{\mu}_{\dot\alpha}(x)j_{\beta}(0)\rangle\sim\bar\sigma^{\mu\gamma}_{\ \dot\alpha}\langle
G_{\gamma}(x)j_{\beta}(0)\rangle=\sigma^{\mu}_{\beta \dot\alpha}f_{Gj}(x^2M^2)~,\cr
&\lim_{|x|\rightarrow\infty}\langle
S^{\mu}_{\alpha}(x)j_{\beta}(0)\rangle\sim\sigma^{\mu}_{\alpha\dot\gamma}\langle \bar
G^{\dot\gamma}(x)j_{\beta}(0)\rangle=\sigma^{\mu}_{\alpha\dot\gamma}\bar\sigma^{\nu\dot\gamma}_{\
\beta}\partial_{\nu}f_{\bar Gj}(x^2M^2)~.}}

The free equation of motion of the Goldstino (alternatively, a comparison with~\threeid\ or current conservation) yields $f_{Gj}=0$. In addition, since in \Sjii\ we showed that only the anti-symmetric combination survives we get that $f_{\bar Gj}=0$ as well.
We conclude that neither the Goldstino $G_{\gamma}$ nor $\bar G_{\dot \gamma}$ mix with $j_{\beta}$ and that therefore the IR behavior is free of one particle exchanges at long distances.

We can now go back to the IR behavior of the correlation functions in~\currtwopt. Using the information we have gathered so far, we can immediately say something about the long distance behavior of $\langle J(x)J(0)\rangle$. The fact that $c=0$ in~\twopoint\ implies that $J$ has no overlap with a one pion state. One can use an analogous argument to show that, in the case of a spontaneously broken R-symmetry, $J$ has no overlap with a single R-axion state. Therefore, the $x^{-2}$ term at large $x$ in $\langle J(x)J(0)\rangle$ is absent, or, in other words,
\eqn\JJconnlR{
\langle J(x)J(0)\rangle_{\rm connected}=\CO(x^{-4})~,
}
where the left hand side is the connected part of the two-point function.\foot{There can also be a disconnected contribution given by $\langle J\rangle^2=\zeta^2$. In addition, one can show that the actual decay rate of the correlation function in~\JJconnlR\ is faster than the conservative bound we presented in~\JJconnlR.  }

Next, let us consider the $\langle j_{\alpha}(x)\bar j_{\dot\beta}(0)\rangle$ and $\langle j_{\alpha}(x)j_{\beta}(0)\rangle$ correlators.
In the absence of other massless fermions the only dangerously singular IR behavior of $\langle j_\alpha(x) j_\beta(0)\rangle$ and $\langle j_\alpha(x) \bar j_{\dot\beta}(0)\rangle$ could arise from single Goldstino exchange. We have seen above that this does not happen.

The only remaining correlation function to discuss is the two-point function of the spin-1 current $\langle
j_\mu(x) j_\nu(0)\rangle$. On general grounds, however, this correlator will receive a contribution from an
exchange of a pion at tree level. Indeed, we expect at long distances \eqn\jmujnucorrIR{\eqalign{
\lim_{|x|\rightarrow\infty}\langle j_{\mu}(x)j_{\nu}(0)\rangle\sim \partial_{\mu}\partial_{\nu}\left({1\over
x^2}\right)~. }} Thus, this is the only correlation function in which there is a long distance contribution from
a single particle exchange.

Throughout the analysis we have assumed that the massless particles are a pion, a Goldstino, and possibly an R-axion. To complete our discussion we would like to comment on the case in which there are additional massless 't Hooft fermions. 't Hooft fermions are, of course, charged under some unbroken global symmetries. Since $j_\alpha$ is neutral under all the unbroken global symmetries it can not mix with charged fermions, and therefore these fermions do not lead to single particle exchange diagrams at large separation.

This argument fails for unbroken R-symmetry. 't Hooft fermions with R-charge $-1$ can mix with $j_\alpha$ and lead to a term of the form $\partial(x^{-2})$ in the correlation function $\langle j_\alpha(x)\bar j_{\dot\alpha}(0)\rangle$ at long distances. Of course, an exact unbroken R-symmetry is undesirable phenomenologically, but this possibility should be kept in mind.\foot{A phenomenological scenario with unbroken R-symmetry was suggested in~\FoxBU. The presence of such $R=-1$ massless fermions should play a crucial role in this case. See~\AmigoRC,\BenakliPG\ and references therein for more recent works. }
\newsec{Gauging the Symmetry}

In the previous section we treated the symmetry $\CG$ as global. In this section we would like to weakly gauge it and consider the resulting gauge-mediated soft terms for visible sector fields charged under $\CG$. In the discussion that follows, we will identify $\CG$ with an additional $U(1)$ gauge symmetry of the visible sector. Therefore, our previous emphasis on studying the effects of Higgsing will turn out to be of immediate relevance, since any additional $U(1)$ gauge symmetry of the type we are interested in must be Higgsed above the electroweak scale. Furthermore, we will find that the IR contributions to the two-point functions we discussed in the previous section will play a starring role in our discussion of the soft terms below.

In order to proceed, we must give a prescription for how to weakly gauge $\CG$. The essential technique for carrying out this procedure in the case of un-Higgsed symmetries was laid out in~\MeadeWD\ and follows from the definition of gauge mediation given in the introduction. The basic idea is to couple the corresponding global current superfield, $\CJ$, to a vector superfield, $\CV$, and to work perturbatively in the resulting coupling $g$. The relevant SUSY breaking data of the hidden sector is summarized in the exact one and two-point functions of $\CJ$ discussed in the previous section.

In this section we will study the soft masses induced in the presence of Higgsed symmetries. Since our vacuum by definition has at least two massless
particles, the goldstino and the pion of $U(1)$, we will need to exercise extra care regarding the finiteness and IR safety
of the contributions to the visible soft masses.

For simplicity we assume that the spectrum of $\CH$ has a typical scale $M$ at which the dynamics takes place. In the deep IR,
we find the goldstino, the Goldstone boson, and, potentially, some massless fermions. Since the $U(1)$ is Higgsed at the scale $M$,
the associated vector field will have a mass of the order
$gM$. The mass splittings in the vector superfield, together with the mass splittings in the hidden sector itself, will then generate the soft scalar masses.

To begin our discussion, let us couple the  vector superfield $\CV$ to the hidden sector current superfield, $\CJ$
\eqn\gauging{
\CL_{\rm int}=2g\int d^4\theta\CJ \CV =g(JD-\lambda  j-\bar\lambda\ \bar j-j^{\mu}V_{\mu})~.
}
We have chosen to write the vector superfield in WZ gauge. This choice is natural even if $\CG$ is Higgsed since we are interested in the physics at high scales as well as at low scales.
Integrating out the hidden sector in~\gauging\ leads to the following effective action for the vector multiplet
\eqn\Veff{\eqalign{
\delta\CL_{\rm eff}&=g\xi D+{g^2\over2}\tilde C_0(p^2)D^2-ig^2\tilde C_{1/2}(p^2)\lambda\sigma^{\mu}\partial_{\mu}\bar\lambda-
{g^2\over4}\tilde C_1(p^2)F_{\mu\nu}F^{\mu\nu}\cr&-{g^2\over2}(M\tilde B_{1/2}(p^2)\lambda\lambda+c.c.)...
}}

In the previous section we analyzed the low momentum behavior of the different functions $\tilde C_i,\tilde B$ appearing above.
We concluded that generic SUSY breaking theories lead to
\eqn\funcb{\eqalign{&
\tilde C_0(p^2)=\tilde C_0(0)+\CO(p^2/M^2)~,\cr
&\tilde C_{1/2}(p^2)=c_{1/2}{M^2\over p^2}+\hat{\tilde C}_{1/2}(p^2)~,\cr
&\tilde C_{1}(p^2)=c_1{M^2\over p^2}+\hat{\tilde C}_1(p^2)~,\cr
&\tilde B_{1/2}(p^2)=\tilde B_{1/2}(0)+\CO(p^2/M^2)~.
}}
The functions $\hat{\tilde C}_{1/2}(p^2)$, $\hat{\tilde C}_{1}(p^2)$ are by definition regular at zero momentum. Note that $\tilde C_{0}$ is
a regular function at zero momentum, as follows from our discussion in section 2. As long as the $U(1)$ is spontaneously
broken we know that $c_1\neq0$. On the other hand, $c_{1/2}\neq0$ only when there is an
unbroken R-symmetry and a massless fermion with charge~$-1$ under the R-symmetry.\foot{We assume genericity here. As we remarked in section 2,
specific models, perhaps with multiple sectors, can violate this statement. For example, one can include a separate supersymmetric sector. Since this setup is not
well motivated, we ignore this possibility. All our statements are easily adaptable to more peculiar cases. }

Let us now analyze the effective action~\Veff\ in more detail. We notice that
in the case of interest $c_1\ne0$ so $\CL_{\rm eff}$ contains singular terms. If $c_{1/2}\neq 0$ then the gaugino propagator is also seemingly singular. The appearance of singular terms is due to the fact that
there are massless particles in the hidden sector. However, this framework is still legitimate because we making an expansion in $g\ll1$.

To understand the meaning of the singular terms in~\Veff, we can study various tree level two-point functions of the vector multiplet.
We fix the remaining gauge freedom by choosing a Feynman gauge.\foot{More precisely, we choose the gauge fixing part of the Lagrangian to be
$$ \delta\CL_{\rm gf}=-{1\over2}(1+g^2\tilde C_1(p^2))(\partial_{\mu}A^{\mu})^2~.$$
}
The vector boson propagator is \eqn\Vbosprop{ \langle
V^{\mu}V^{\nu}\rangle={g^{\mu\nu}\over p^2(1+g^2\tilde C_1(p^2))}~.} Now, expanding the denominator of
\Vbosprop, we find $p^2+g^2c_1M^2+\CO(p^2g^2)$. We therefore see that the vector boson
acquires a mass via the Higgs mechanism \eqn\vbosmass{ m_V^2=g^2c_1M^2+\CO(g^4)~. } Similarly, we can compute the gaugino
propagators \eqn\ggbar{\eqalign{ &\langle \lambda_{\alpha}\bar\lambda_{\dot\beta}\rangle
=-{p_{\mu}\sigma^{\mu}_{\alpha\dot\beta}\over p^2(1+g^2\tilde C_{1/2}(p^2))+{g^4|M\tilde B_{1/2}(p^2)|^2\over
1+g^2\tilde C_{1/2}(p^2)}}~, \cr&\langle\lambda_{\alpha}\lambda_{\beta}\rangle
=-\epsilon_{\alpha\beta}{g^2M\tilde B_{1/2}(p^2)\over p^2(1+g^2\tilde C_{1/2}(p^2))^2+g^4|M\tilde
B_{1/2}(p^2)|^2}~. }} Expanding the denominator to $\CO(g^2)$, we find
$p^2+g^2c_{1/2}M^2+\CO(p^2g^2,g^4)$.
In particular, if $c_{1/2}\neq 0$ (as can happen if the R-symmetry is unbroken and there is a massless fermion with R-charge $-1$), the gaugino is mostly
a Dirac particle with mass
\eqn\diracg{
m^2_{\lambda}=g^2c_{1/2}M^2+\CO(g^4)~.}

The situation in which $c_{1/2}=0$ is slightly more subtle. In the case that R-symetry is unbroken, $\tilde B_{1/2}=0$. From the first line of \ggbar, we then see that $\langle \lambda_\alpha\bar\lambda_{\dot\beta}\rangle$ has a zero momentum pole and so the $U(1)$ gaugino remains massless.

If, on the other hand, $c_{1/2}=0$ and R-symmetry is broken, the gaugino must be a Majorana particle whose mass is expected to scale as $g^2M$. Note that in theories with perturbative control, this mass is expected to arise without a $1/16\pi^2$ factor. The reason for this behavior is that the current $j_\alpha$ generally mixes directly with massive hidden sector fermions.

We conclude that in generic theories of the type we are interested in, there is a hierarchical cascade of scales: The hidden sector dynamics occurs at the scale $M$. The mass of the $U(1)$ gauge boson is $gM$, and the mass of the $U(1)$ gaugino is $g^2M$. To complete this picture, we will discuss the contributions to the soft scalar masses in the next subsection.

As a final note before proceeding, let us comment on the expansion in $g$. The effective theory~\Veff\ is valid up to but not including corrections of order
$\CO(g^4)$. Our expressions for the propagators~\Vbosprop, \ggbar\ are meaningful within this perturbation
theory. For instance, the denominator of~\Vbosprop\ is expected to have $g^4$ corrections which we can safely
ignore for our purposes here.

\subsec{Scalar Masses}

From the effective theory~\Veff\ we can evaluate the soft scalar masses. If there is no messenger parity we expect $\zeta\neq0$ and the contribution to the scalar masses to arise at tree-level. The result would be
$\delta m^2_{soft}\sim g^2\zeta$. The more interesting option is when $\zeta=0$ due to unbroken messenger parity (or when $\zeta$ is very small due to approximate messenger parity).\foot{The former option is not appealing phenomenologically as it would require an unnatural choice of the gauge coupling to make $\delta m^2_{soft}$ comparable to or smaller than the usual gauge mediated contributions.} If this is the case, the leading source for soft scalar masses comes from loops involving the effective theory~\Veff. Our purpose in this section is to analyze the loop integrals which determine the leading contribution to the soft scalar masses. Thus, in the rest of this subsection, we assume that $\zeta=0$.

Since we wish to extract the leading order result in a systematic expansion of the scalar masses in terms of the coupling $g$, we have to pick only the relevant piece of the propagators~\Vbosprop,~\ggbar. Due to the zero-momentum singularity in $\tilde C_{1}$, we cannot naively expand the propagator~\Vbosprop\ in $g$.
Summing the one-loop integrals in the effective theory~\Veff\ we get the following expression for the soft scalar masses
\eqn\correction{ m^2_{soft}=-q^2g^2\int{d^4p\over(2\pi)^4}\sum_i{(-1)^{2i+1}N_i\over
p^2\left(1+g^2\tilde C_i\right)}~,} where $N_0=1, N_{1/2}=4, N_1=3$ and $q$ is the $U(1)$ charge of the corresponding sparticle. If the functions $\tilde C_i$ were all nonsingular, we could have expanded in $g$ and rediscovered the familiar expression of GGM. In our case, although $\tilde C_{0}$ is generically a well behaved function around zero momentum, $\tilde C_{1/2}$ and $\tilde C_1$ are potentially singular. Let us assume for simplicity that
R-symmetry is broken (spontaneously or explicitly). Then, it follows from our discussion in section~2 that $\tilde C_{1/2}$ is regular at zero momentum. On the other hand, $\tilde C_1$ is necessarily singular at zero momentum.

Let us focus on the integral involving $\tilde C_1$
\eqn\inte{3q^2g^2\int {d^4p\over (2\pi)^4}{1\over p^2\left(1+g^2\tilde C_1(p^2)\right)}~.}
At very small $p$, $\tilde C_1(p^2)$ can be approximated by $c_1/p^2$. Thus, the contribution that we get from this region is
\eqn\inte{3q^2g^2\int {d^4p\over (2\pi)^4}{1\over p^2+g^2c_1}~.}
This integral is convergent at small momentum, and if we integrate from $p=0$ to $|p|=\Lambda$ we get that the $g$ dependence is $g^4c_1\log\left(g^2c_1\over \Lambda^2+g^2c_1\right)$. The approximation of $\tilde C_1(p^2)$ by $c_1/p^2$ is valid for momentum much smaller than the typical scale $M$. Thus, we can still take the cutoff of the integral to be
$\Lambda\gg g^2c_1$. We see that the integral becomes $g^4c_1\log\left(g^2c_1\over \Lambda^2\right)$. This dependence on $g$, which does not arise in the usual setup of GGM, is a clear sign of the Higgsed case. It reflects the bad behavior of $\tilde C_1$ at low momentum.

As we see explicitly in our above discussion, there is a logarithmic sensitivity to the cutoff scale $\Lambda$. This expression is regulated by contributions to the various $\tilde C_i$ from states at the scale $M$. These corrections remove the logarithmic divergence, but they are analytic in $g$ and therefore cannot cancel the $\log(g^2)$ piece.

The coefficient of $g^4\log (g^2)$ depends only on $c_1$ which is directly related to the mass of the vector field and in this sense is universal. We can easily read the precise coefficient from~\inte\ with the result
\eqn\universal{\delta m^2_{soft}={3q^2g^4\over 16\pi^2}\log(g^2)c_1+\CO(g^4)={3q^2g^2\over 16\pi^2}\log(g^2)m_V^2+\CO(g^4)~.}
The $\CO(g^4)$ corrections to the formula~\universal\ stand for terms that do not contain a logarithm of $g^2$ and are therefore parameterically suppressed. We see that the sign of the leading term in the soft scalar mass~\universal\ is negative and, as mentioned above, depends only on the scale of Higgsing.

Another important difference from the un-Higgsed case is that the leading term in~\universal\ has only one factor of $16\pi^2$ in the denominator, while in the un-Higgsed case there are two such factors. Moreover, since the logarithmic sensitivity we have seen above also comes with one $16\pi^2$ in the denominator, the threshold corrections which remove the logarithm will have one factor of $16\pi^2$.

Finally, there are threshold corrections at the scale $M$ which are genuine two-loop diagrams. These resemble the usual gauge mediated contributions and have a similar dependence on $g$ and two $16\pi^2$ factors. These contributions are negligible compared to the universal term~\universal\ and the threshold corrections mentioned above.

The basic physical reason for this structure can be understood heuristically as follows: There are two sources for non-supersymmetric multiplets.
One is the gauge multiplet and the other is the hidden sector itself.
The universal result~\universal\ can be derived from the one-loop diagrams associated with the non-supersymmetric vector superfield, as in Figure~1.
Other one-loop diagrams containing hidden sector fields will have to come in at the scale $M$ to cancel the logarithmic divergence. Finally there are genuine two-loop diagrams involving hidden sector fields which give rise to contributions that are similar to the ones found in the usual gauge mediated scenarios. In our case these contributions are negligible.

\ifig\neffrbfig{The most general set of diagrams contributing to the soft mass of a scalar $Q$ has as the leading contribution an exchange of the $U(1)$ pion. This can also be thought of as a diagram with an insertion of the gauge field mass term. The logarithmic divergence is canceled by massive particles in the hidden sector. The remaining signature of the leading diagram is a nonanalytic contribution $g^4\log(g^2)$ to the soft mass of $Q$.}{\epsfxsize=0.8\hsize\epsfbox{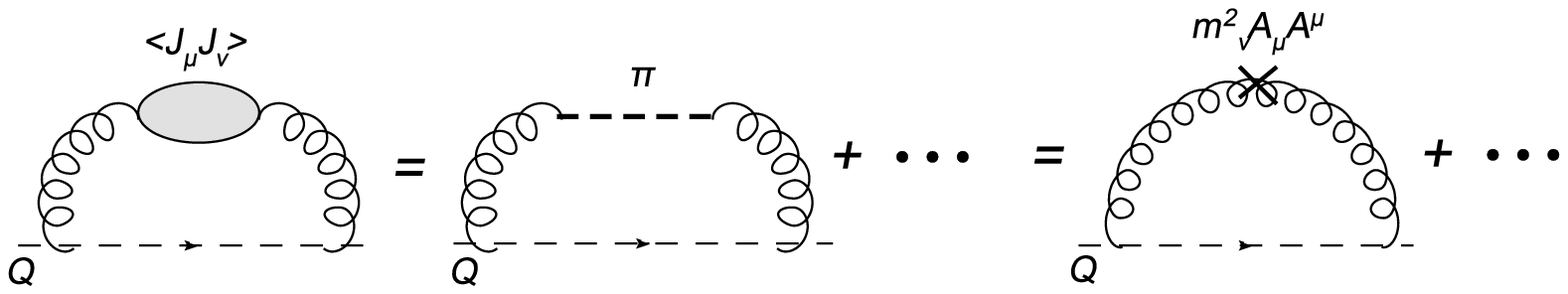}}

Our results, including the universal contribution~\universal\ and the existence of hidden sector threshold corrections which come with a single $16\pi^2$ factor, are completely general.

We finish this section by commenting that if, for instance, $\tilde C_{1/2}$ has a singularity at zero momentum the discussion above has to be modified in a straightforward way. We have already emphasized that this looks highly non-generic and phenomenologically undesirable. So, we view the prediction~\universal\ and the various consequences we have discussed as robust.

\newsec{Toy Model}

We would like to illustrate some of the points in the previous section in a specific weakly coupled example that involves spontaneous supersymmetry and global symmetry breaking.
To that end, consider the following O'Raifeartaigh-like model
\eqn\supe{{ W}=X(f-\phi_+\phi_-)+m \ \eta_+\phi_-+m \ \phi_+\eta_-~.}
This theory has a global $U(1)$ symmetry and a $U(1)_R$ symmetry under which the fields have the following charges
\eqn\tableone{\matrix{& U(1) & U(1)_R \cr & \cr  \phi_+ & 1 & 0 \cr  \phi_- & -1 & 0 \cr  \eta_+ & 1 & 2 \cr  \eta_- & -1 & 2 \cr  X & 0 & 2    }}
In addition, the theory has a messenger parity symmetry under which $\phi_+\rightarrow\phi_-$, $\eta_+\rightarrow\eta_-$.

We can take all the parameters of the superpotential to be real and positive without loss of generality. We will be interested in the regime $f>m^2$ where the $U(1)$ symmetry is spontaneously broken.

To find the minimum of the potential we  can set $F_{\phi_+,\phi_-}=0$. This gives the following relations
\eqn\relations{
m \eta_+= X\phi_+~,\qquad m \eta_-= X\phi_-~.}
The remaining terms in the scalar potential have a SUSY-breaking minimum (which is also the global minimum of the potential) with
\eqn\solution{
\phi_+=\phi_-^*~,\qquad |\phi_+|^2=f-m^2.
}
This describes the $S^1$ associated with the spontaneous breaking of  the $U(1)$ symmetry.
We conclude that the lowest lying SUSY-breaking solution is part of a classical moduli space consisting of an $S^1$ fibered over the complex plane described in~\relations.\foot{Complex flat directions are ubiquitous in such theories, see~\RayWK,\KomargodskiJF.} A convenient coordinate on this complex plane is the expectation value of $X$.
The vacuum energy is
$V_0=m^2(2f-m^2)$.\foot{At $f=m^2$ there is a transition to a symmetry restoring vacuum in which $\langle\phi_+\rangle=\langle\phi_-\rangle=0$. The vacuum energy changes continuously across this transition.}

The complex flat direction parameterized by~\relations\ is lifted at one-loop. As a result, $X$ is stabilized at the origin, $\langle X\rangle=0$. Consequently, $\eta_{\pm}$ are also stabilized at the origin.\foot{Eventually, when we gauge the $U(1)$ symmetry, this is assumed to be weak enough such that the vacuum is almost not shifted.}  This means that the $U(1)_R$ symmetry remains unbroken. Moreover, messenger parity symmetry also remains unbroken in this vacuum.

The unbroken $U(1)_R$ symmetry renders this model unrealistic but this problem is very easy to take care of by making the model slightly more complicated. Our purpose here is merely to illustrate some of the general results we discussed in the previous section, so for this sake the simple model~\supe\ suffices.

In accord with our general discussion, $c_0=0$ in this theory and $c_1$ is given by the decay constant of the pion, $c_1\sim f-m^2$. It is easy to check that besides the Goldstino there are no other massless fermions and so $c_{1/2}=0$.\foot{In particular, although R-symmetry is unbroken, there is no massless fermion with R-charge~$-1$.}

The current superfield of the $U(1)$ symmetry is
\eqn\cj{\eqalign{
&\CJ=\CJ_{0}+\CJ_{2}~,\cr&\CJ_0=\phi_+^{\dagger}\phi_+-\phi_-^{\dagger}\phi_-~,\cr&\CJ_{2}=\eta_+^{\dagger}\eta_+-\eta_-^{\dagger}\eta_-
~.}}
We have separated the current superfield into two pieces $\CJ_0$ and $\CJ_2$ with indices corresponding to the  R-charges of the various chiral fields appearing in \cj. The unbroken R-symmetry guarantees that $\CJ_0$ and $\CJ_2$ do not mix to the order we are interested in.

The general formula for the scalar masses is given by
\eqn\scalarmasstoyii{\eqalign{
&m_{soft}^2=q^2g^4{f-m^2\over 4\pi^2}\Big(4\log2+3\log{2g^2(f-m^2)\over
m^2}\Big)+\CO\left({g^4\over (16\pi^2)^2}\right) ~.}}
The first term in~\scalarmasstoyii\ corresponds to threshold corrections from the hidden sector which are responsible for the cancelation of the logarithmic divergence we described in the previous section. As the general discussion in section~3 implies, these threshold corrections arise from one-loop diagrams. In the language of our specific weakly coupled hidden sector, the threshold corrections which remove the logarithm come from particles in the ``pion supermultiplet." In general these corrections are model dependent. The second term in~\scalarmasstoyii\ contains the universal term we derived in~\universal. It comes from the one-loop diagram involving the massive vector field. For small enough gauge coupling $g$ this is the leading term and it contributes negatively to the soft scalar masses.

\newsec{Discussion and Phenomenological Applications}
As we have seen in our discussion, under very general assumptions, the dynamics of the hidden sector occurs at
the scale $M$, the $U(1)$ gauge field acquires a mass of order $gM$, the gaugino of this additional $U(1)$ acquires a mass of order
$g^2M$, and the visible sector scalars acquire a new contribution to their mass-squared of order $\left(g^2M/4\pi\right)^2$ (where we drop the logarithm for
simplicity). It is also rather straightforward to determine the scale of the A-terms that are induced in our
setup. Indeed, the A-terms are generated at one-loop and are of the form \eqn\aterm{ A\sim{g^4\over 16\pi^2}M~. } The cascade of scales we encounter in this class of models is represented in Figure 2.

\ifig\neffrbfig{The general cascade of scales in a theory where the dynamics of the hidden sector happens at a scale $M$. As we have explained, the precise formulae may also include logarithms of $g$ and a specific dependence on the charges of the visible particles under the $U(1)$ symmetry.}{\epsfxsize=0.5\hsize\epsfbox{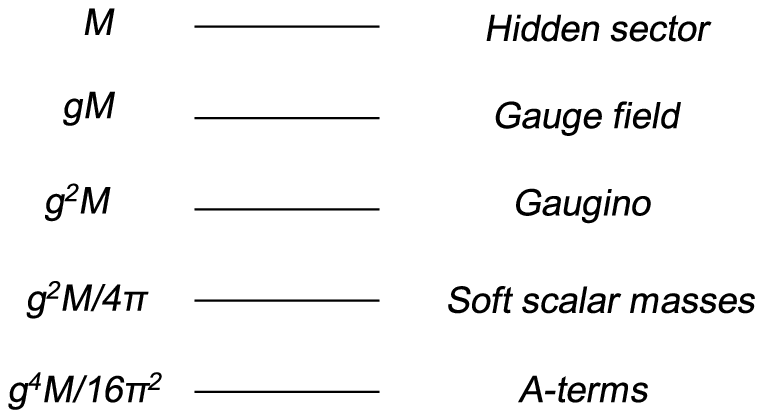}}

Our entire discussion fits naturally into the scheme of low-scale SUSY breaking. Taking $g\sim10^{-1}$ (which is
of the same order as other gauge couplings in the SSM) and assuming the hidden sector scale is $M\sim 100 \ {\rm
TeV}$, we find that the gauge boson has a mass of order 10~TeV, the gaugino of the new $U(1)$ has a mass of order 1~TeV, the
scalars receive a contribution to their soft mass-squared of order $10^4~{\rm GeV^2}$, and the A-terms are of order 100~MeV. As usual in gauge mediation,
the A-terms we generate are parameterically small. We see that this natural choice of parameters leads to
sizeable modifications of the soft scalar masses. Therefore, the presence of additional (Higgsed) gauge
symmetries can significantly alter the spectrum of the sparticles.

The fact that the leading contribution is fixed and negative provides a potential mechanism for lowering the scalar masses in gauge mediation, which is phenomenologically desirable. On the other hand, the fact that the scalar masses are negative forces the new contributions to be tightly bounded, otherwise some of the sparticles may become tachyonic.

It is also interesting to note that these contributions from
additional $U(1)$ symmetries  violate the sum rules of gauge
mediation. If there is only one additional $U(1)$ symmetry, one sum
rule always remains.\foot{As noted in the introduction, this
statement is true if the correlation functions
$\langle\CJ(x)\CJ_Y(0)\rangle$ are small due to, e.g., exact or
approximate messenger parity, or either of the gauge symmetries
being embedded in a non-Abelian structure.} The form of the sum rule
depends on the charges of the visible particles under the $U(1)$
symmetry. For example, if we choose the additional $U(1)$ to be
$U(1)_{B-L}$ then we find that the only remaining sum rule is $\Tr(Y
m^2)=0$.\foot{For completeness, we quote here the general remaining
sum rule
\eqn\sumrule{\eqalign{&m_Q^2(3q_U^2-3q_D^2-q_E^2)+m_U^2(-3q_Q^2+3q_D^2+3q_L^2-q_E^2)+m_D^2(3q_Q^2-3q_U^2-3q_L^2+2q_E^2)\cr&+m_L^2(-3q_U^2+3q_D^2+q_E^2)+m_E^2(q_Q^2+q_U^2-2q_D^2-q_L^2)=0~.}}
The $q_i$ represent the charges of the corresponding matter fields under the Higgsed $U(1)$ symmetry. This result is in agreement with~\LuoKF.} Since the term violating the $\Tr((B-L) m^2)$ sum rule has a fixed sign at leading order in $g$ this implies that (at leading order)  \eqn\pred{\Tr\left((B-L)m^2\right)>0~.}

In our estimates above we have assumed that the vacuum respects messenger parity. For completeness we would like
to mention what happens if this assumption is relaxed. It is still true that the gauge fields have mass of the
order $gM$. The $U(1)$ gauginos have mass $gM$ or $g^2M$ depending on whether $c_{1/2}\neq 0$ or $c_{1/2}=0$,
respectively. Due to the fact that we do not have messenger parity, a $D$-term is generically induced. Therefore
the scalar masses typically arise at order $gM$. This scenario, where everything is essentially induced at
tree-level, does not seem appealing because it requires rather small values of the gauge coupling constant. One
should bear in mind that more elaborate models with more scales may lead to different behavior. Here we just
present the tools and methods to address these questions, emphasizing the universal results.

Finally, let us discuss some open questions. First, it would be nice to build explicit calculable models of
the kind we discussed---including the full SSM gauge group and a broken R-symmetry---and study the masses of the sparticles explicitly. It would also be interesting to embed such models in string theory and study how generically scenarios like the one we have described occur. Perhaps the significant, negative, contributions to the sfermion mass squareds that we have identified could constrain the string constructions in some interesting ways.

Of course, we would also like to study different classes of models than the ones we have studied here. Notice that the models we discussed, with parametrically small gauge coupling, fit naturally into low scale gauge mediation. However, there is another class of models in which the parameter of SUSY
breaking $F/M^2\ll 1$ is the smallest quantity in the theory, not the gauge coupling. These models appear naturally in high scale gauge mediation. It would be interesting to
understand models of this type in some generality and compare them to models of the kind we study. A closely related
problem is to study theories where the gauging of the global symmetry of the hidden sector has important effects
on the dynamics.\foot{A well known example is the inverted hierarchy~\WittenKV.} The expansion in $g$ is perplexing
in these cases and it would be useful to have a framework for such models.\foot{We should also note that theories of $Z'$ mediation \LangackerIP, \LangackerAC\ lie outside the class of theories we have considered in this paper. Indeed, in $Z'$ mediation, the breaking of the additional $U(1)$ symmetry arises through couplings to the SSM Higgs fields.}
\bigskip
\bigskip
\centerline{\bf Acknowledgements}

We would like to thank N.~Arkani-Hamed, M.~Dine, G.~Festuccia, N.~Seiberg, and D.~Shih for interesting discussions. The work of MB was supported in part by NSF grant PHY-0756966 and the ERC Advanced Grant 226371. The work of ZK was supported in part by NSF grant PHY-0503584. Any opinions, findings, and conclusions or recommendations expressed in this material are those of the author(s) and do not necessarily reflect the views of the funding agencies.

\listrefs
\end